\documentclass[aps,pre,groupedaddress,twocolumn,showpacs]{revtex4-1}

\renewcommand{\vec}[1]{\ensuremath{\underline{#1}}}

\usepackage{amsmath}
\usepackage{graphicx}

\begin{document}

\title{Jamming and Attraction of Interacting Run-and-Tumble Random Walkers}

\author{A. B. Slowman}
\author{M. R. Evans}
\author{R. A. Blythe}
\affiliation{SUPA, School of Physics and Astronomy, University of Edinburgh, Peter Guthrie Tait Road, Edinburgh EH9 3FD, UK}

\date{5th May 2016}

\begin{abstract}
We study a model of bacterial dynamics where two interacting random walkers perform run-and-tumble motion on a one-dimensional lattice under mutual exclusion and find an exact expression for the probability distribution in the steady state. This stationary distribution has a rich structure comprising three components: a jammed component, where the particles are adjacent and block each other; an attractive component, where the probability distribution for the distance between particles decays exponentially; and an extended component in which the distance between particles is uniformly distributed. The attraction between the particles is sufficiently strong that even in the limit where continuous space is recovered for a finite system, the two walkers spend a finite fraction of time in a jammed configuration. Our results potentially provide a route to understanding the motility-induced phase separation characteristic of active matter from a microscopic perspective.
\end{abstract}

\pacs{05.40.-a; 87.10.Ca; 87.10.Mn}


\maketitle

Self-propelled particles consume energy in order to generate persistent motion and typically self-organize into complex structures \cite{Toner2005}. These particles may be naturally occurring, for example, birds that flock \cite{Ballerini2008}, or synthetic, such as photoactivated colloids that form `living crystals' \cite{Palacci2013}. It has become apparent that the physics of such active constituents may be far richer than traditional passive, equilibrium matter.
 
The key distinction between passive and active particles at the microscopic level is that the equations of motion for the latter break time-reversal symmetry (also known as detailed balance). For example, the continual consumption of energy implies that individual collisions do not need to conserve energy or momentum.  At the macroscopic scale, a robust finding is that self-propelled particles exhibit motility-induced phase separation \cite{Cates2015}: that is, a tendency to cluster as a consequence of the particle velocity decreasing as the local  particle density increases. The propensity for clusters to form is of great interest from a fundamental perspective, and has given rise to  a variety of theoretical and computational studies  \cite{GRedner2013,Bricard2013,Elgeti2015,Stenhammar2013,Wittkowski2014,Speck2014,Detch15}. Moreover, clustering may have practical implications: for example, bacteria are commonly found in aggregates called biofilms which are important sources of human infection \cite{Costerton1999Sci,Davies2003} and contamination in the food industry \cite{Simoes2010}. 

Although various theoretical approaches have successfully reproduced some of the macroscopic properties of clustering, most insights have arisen by coarse-graining over microscopic degrees of freedom to a greater or lesser degree (see e.g.~\cite{Schimansky1995,TonerTu1995,Tailleur2008,Fily2012} and also \cite{Toner2005,BenJacob2010,Marchetti2013,Cates2015} for reviews). This coarse-graining step leaves one unable to pinpoint the precise origin of these phenomena. Specifically, although self-propulsion must mediate an effective attraction between otherwise repulsive particles \cite{Hagan2016Em}, no systematic method for determining the exact form of this emergent attraction from the underlying microscopic dynamics exists. Such a method would pave the way towards a deeper understanding of the mechanism behind motility-induced phase separation. Recent theoretical investigations have used a microscopic approach to generate effective interactions, but they have been of an approximate nature \cite{Farage2015,Maggi2015}. It thus remains of paramount importance to establish exact results that shed light on the path from the microscopic breaking of detailed balance to the emergence of effective attractions.

In this work, we determine the exact analytical form of the effective pair potential that emerges between a pair of self-propelled particles undergoing the run-and-tumble dynamics that characterizes certain bacterial species (notably \textit{Escherichia coli} \cite{Berg1972,Berg2004}).  In its most idealized form \cite{Schnitzer1993},  run-and-tumble  motion consists of a series of straight-line runs at velocity $v$, interspersed by tumble events that occur as a Poisson process with rate $\alpha$ and which instantly randomize a particle's direction of motion. Although in general the run velocity $v$ may depend on the local density of bacteria or the concentration of various chemical species in the environment \cite{Schnitzer1993,deGennes2004,Kafri2008,Tailleur2008}, and time is spent tumbling without moving \cite{Berg1972}, we consider the simplest case where the velocity $v$ is constant and the motion is in one dimension. Thus when a tumble occurs with rate $\alpha$, a velocity $+v$ or $-v$ is immediately adopted with equal probability.

We also introduce a hard-core exclusion interaction between the particles: when two particles collide, this interaction causes the particles to remain stationary until one of them reverses its velocity (occurring at rate $\omega=\alpha/2$). It is this specific aspect of the dynamics that breaks detailed balance: energy is not conserved in these collisions.

Our main result is an exact expression for the steady-state probability distribution of this pair of run-and-tumble particles on a periodic lattice, which we can then interpret as an effective pair potential. The distribution has a surprisingly rich structure, and comprises a \emph{jammed} component in which the particles are facing each other on neighboring lattice sites, an \emph{attractive} component characterized by an exponential decay over a finite separation length and an \emph{extended} component in which all microscopic configurations are equally likely. Most remarkably, in a system of finite length, the particles spend a finite fraction of their time in a jammed configuration even if the lattice spacing of the discrete model is taken to zero, in which limit the lattice model recovers run-and-tumble dynamics in continuous space and time. Moreover, the dynamics in the steady state exhibit some intriguing first-passage properties, extending what has been established for individual non-interacting run-and-tumble particles \cite{Angelani2014,Angelani2015}.

\begin{figure}
  \centering
    \includegraphics[width=0.66\linewidth]{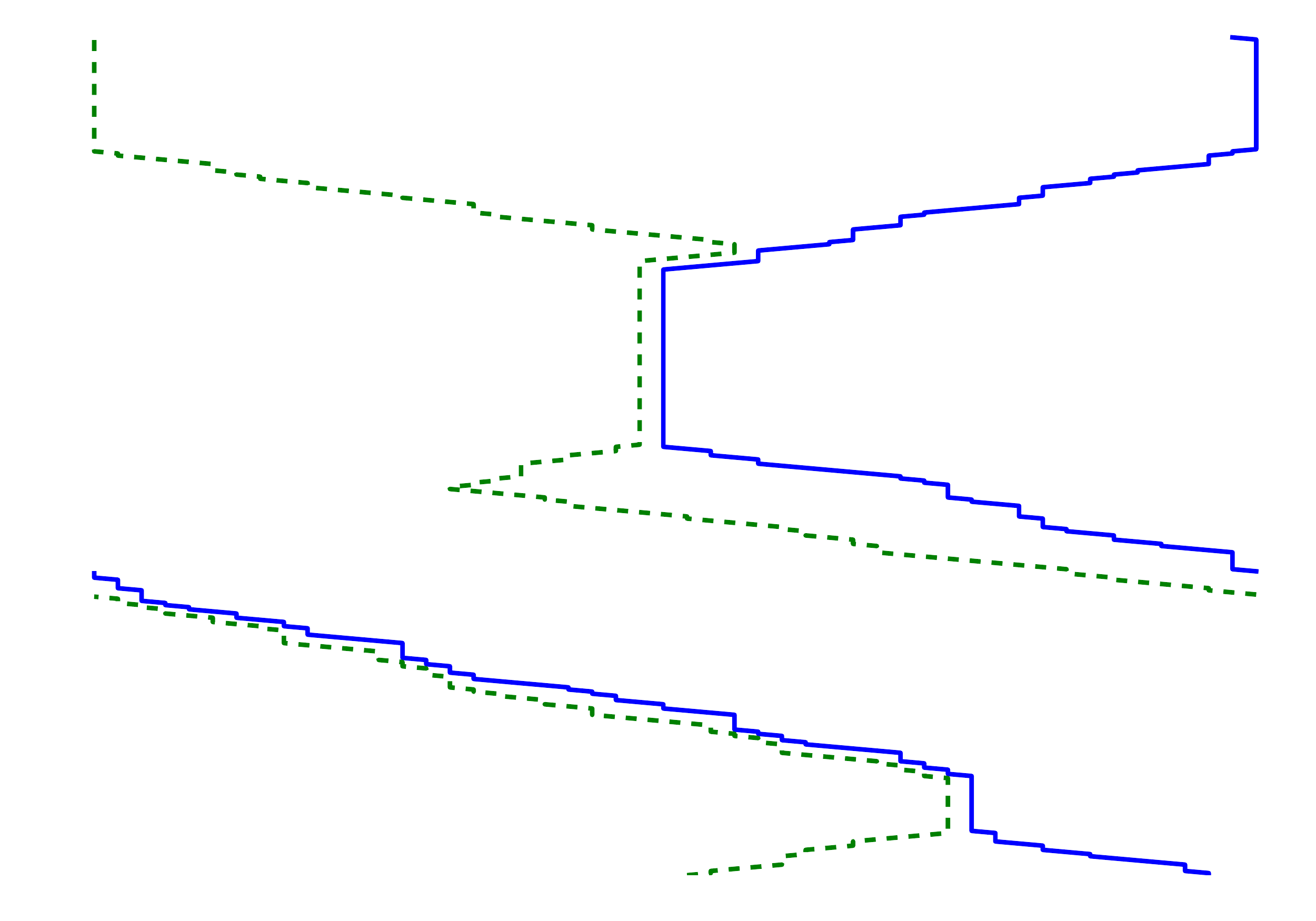}
    \caption{Simulation of Model System: A space-time plot (time in the vertical direction) of a simulation of two run-and-tumble random particles on a one-dimensional ring of $100$ sites in the low tumble-rate regime with particles reversing their direction after traversing $100$ lattice sites on average. The full and dotted trajectories each represent an individual particle.}
    \label{fig:lowtum}
\end{figure}

Let us define our lattice-based model of two run-and-tumble particles in one dimension (see \cite{Thompson2011,Soto2014} for related models).  The particles occupy sites of a periodic one-dimensional lattice of $L$ sites and each has an orientation $\sigma_i=\pm$ indicating its direction of motion. Due to the translational invariance of the system, a microscopic configuration is fully specified by $1\le n < L$, the distance between the two particles in units of the lattice spacing, and the two particle velocities, $\sigma_1$ and $\sigma_2$. A right-moving particle ($\sigma_i=+$) hops one site to the right with rate $\gamma$; likewise, a left-moving particle ($\sigma_i=-$) hops with rate $\gamma$ to the left. The exception is when the target site is occupied by another particle, in which case hopping is not allowed: this implements the hard-core exclusion interaction. Particles may also reverse their velocity at rate $\omega=\alpha/2$, where $\alpha$ is the tumbling rate described above.  By rescaling time, we can take $\gamma=1$ without loss of generality. Fig.~\ref{fig:lowtum} illustrates the two-particle dynamics for the case where $\omega\ll\gamma$.

We now present exact expressions for  the steady state probability $P_{\sigma_1\,\sigma_2} (n)$ of finding
the two particles with velocities $\sigma_1$,$\sigma_2$ and separated by $n$ sites. These read
\begin{align}
P_{++}(n) &=  \frac{1}{Z} \left[p(z)(z^{n}+z^{L-n}) + q(z) \right] \label{pp} \\
P_{+-}(n)  &=  \frac{1}{Z}\left [p'(z)(z^{n}-z^{L-n}) +  q(z) +\delta_{n,1}\Delta(z) \right] \label{pm}
\end{align}
where
\begin{align}
z &= 1+\omega - \sqrt{\omega(2+\omega)} \label{z} \\
p(z) &= 1-z^{2} \label{pz} \\
p'(z) &= \frac{1-z}{1+z}p(z) = (1-z)^{2} \label{ppz} \\
q(z) &= (1-z)^{2}(1-z^{L}) \\
\Delta(z) &= 2(1+z)(z-z^{L}) \label{Delta} \\
Z &= 4 [\Delta(z) + (L-1)q(z) ] \;. \label{Z}
\end{align}
The symmetries of the model imply that $P_{++}(n) = P_{--}(n)$ and $P_{+-}(n) = P_{-+}(L-n)$, leaving only $P_{++}(n)$ and $P_{+-}(n)$ independent. These exact expressions are obtained by
solving the master equation for the stationary probability 
distribution using a generating function approach (see below and Appendix~\ref{app:gf}).  Note that the key parameter $z$ lies in the range $0<z<1$, hence $p(z)$, $p'(z)$, $q(z)$ and $\Delta(z)$ are all positive.

Equations (\ref{pp}) and (\ref{pm}) reveal that the  stationary distribution  is a sum of three distinct components
which we now explicitly identify. At large separations $n,\,L-n\gg1$, we have a  uniform particle distribution
$\propto q(z)$, independent of $n$ as for regular diffusion.  This component of the distribution fills the whole of phase space, and we refer to it as  extended.  At intermediate separations, the  probability distribution for the separation between particles decays exponentially as  $z^n$ with a characteristic lengthscale $\xi=1/|\ln(z)|$.
By analogy with quantum mechanical wavefunctions with exponentially-decaying amplitudes, 
we can think of this attractive component as a bound state.  Finally, there is a contribution from the jammed configurations that have particles facing each other on adjacent sites ($n=1$).  

\begin{figure}
  \centering
    \includegraphics[width=0.8\linewidth]{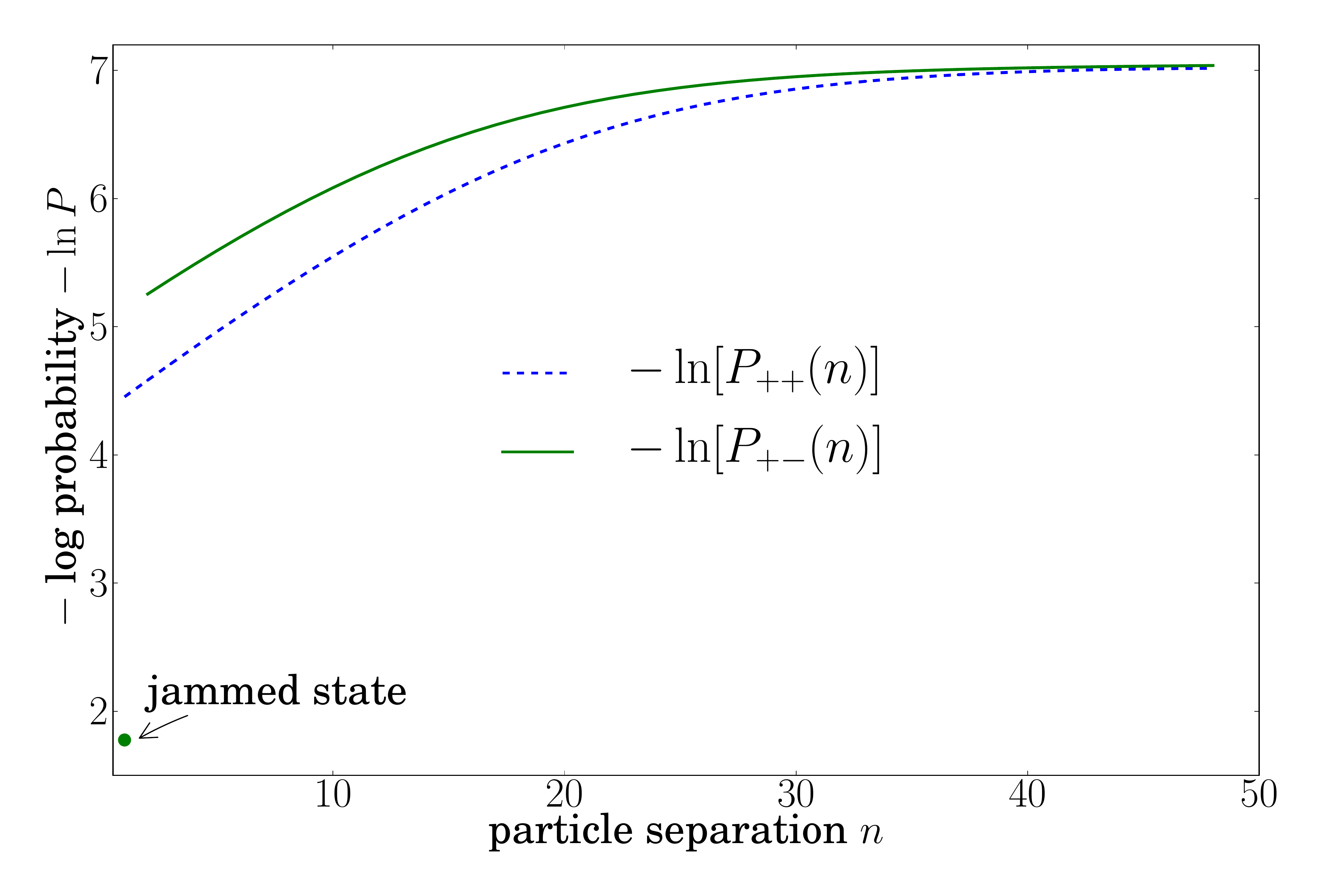}
    \caption{\label{fig:dists} Effective pair potentials, defined by the logarithms of the probability distributions, $P_{++}(n)$ and $P_{+-}(n)$, for the case of $L=100$ lattice sites and velocity reversal rate $\omega = 0.01$. These distributions have three components: jammed (indicated), attractive (linear piece at intermediate separations) and extended (constant piece at large separations).}
\end{figure}

Although the steady state is inherently nonequilibrium, we may nevertheless recast Eqs.~(\ref{pp}) and (\ref{pm}) in the form of effective pair potentials $V_{\sigma_1\sigma_2}(n)=-\ln P_{\sigma_1\sigma_2}(n)$ by analogy with the Boltzmann distribution $P \propto {\rm e}^{-V}$: these are plotted in Fig.~\ref{fig:dists}.  Three distinct  pieces of the potentials
corresponding to the three components of the particle distribution are evident.
At large separations, $n,\,L-n\gg1$, the effective potentials are constant.   At intermediate separations, the potentials are linear and attractive.
Finally, there is a nearest-neighbour ($n=1$) delta function attractive potential. This attraction is very strong when the reversal rate $\omega$ is small.

The origin and physics of the different components of the stationary distribution can be understood from limiting cases. When velocity reversal is rapid, $\omega\to\infty$, we anticipate that standard diffusion should be recovered, as memory of a particle's velocity is erased between each hop. Summing over all four velocity states we obtain the total probability that the two particles are a distance $n$ apart, which in the limit $\omega \gg 1$ becomes
\begin{equation}
P(n) \sim \frac{1}{L-1} \left[ 1 + \frac{1}{2\omega} \left( \delta_{n,1} + \delta_{n,L-1} - {\textstyle\frac{2}{L-1}} \right)  + O\left({\textstyle\frac{1}{\omega^2}}\right) \right] .
\end{equation}
At leading order, only the extended component survives, and we  thus identify repeated velocity reversal as the physical origin of this contribution to the stationary distribution. The jammed component provides the leading correction, whilst the attractive component does not enter at  order  $O(1/\omega)$. 

For the opposite limit, $\omega\to0$,  
the limiting forms of (\ref{pp}) and (\ref{pm}) are
\begin{equation}
P_{++}(n) = \frac{1}{4(L-1)} \quad \mbox{and} \quad
P_{+-}(n) = \frac{1}{4} \delta_{n,1} \;,
\label{Ptum0}
\end{equation}
with corrections  of order $L\omega$, implying that expressions (\ref{Ptum0}) are valid when $\omega \ll 1/L$. In this regime, particles hop many times between velocity reversals, and so in this limit we expect the stationary distribution in each velocity sector to approximate that which would be reached in the absence of any tumbling. For the case where particles are approaching ($+-$), the particles quickly (on the timescale of tumbling) reach the jammed configuration, $n=1$. When they exit this state into one where both particles have the same velocity (e.g., $++$), fluctuations in the distance traveled by each particle, generated by the stochastic particle hopping dynamics, cause the distribution of the relative coordinate to broaden. When this tumble rate is low, the distribution broadens to fill the entire system, thereby generating a uniform distribution, but one that is crucially distinct from the extended component that arises from velocity reversals. This picture of the dynamics is corroborated by  the space-time plots shown in Fig.~\ref{fig:lowtum}. At higher tumble rates, the broadening of the distribution is curtailed on the timescale of tumbling, and is later restarted from the jammed configuration $n=1$. This process is similar to that of diffusion (here, of the particle separation) with stochastic resetting (to the jammed configuration), which generates the exponentially-decaying attractive component of the distribution \cite{Evans2011}. One can thus think of this component as an echo of the jammed configuration.

Finally, and most interestingly, we examine the scaling limit $\omega \to 0$, $L \to \infty$ with $\omega L$ held fixed, in which run-and-tumble dynamics in continuous space and time is recovered. To see why, we introduce the physical system size $\ell$ and reinstate the run rate $\gamma$ which had previously set the unit of time. Then, the mean run velocity is $v = \gamma\ell/L$, and the velocity reversal rate $\omega = \alpha/2\gamma = \alpha\ell/2Lv$, where $\alpha$ is the tumble rate described in the introduction. Substituting into (\ref{pp}) and (\ref{pm}), and introducing the continuous spatial separation $x = n \ell/L$, yields the exact  expressions
\begin{align}
P_{++}(x) &= \frac{\alpha + 2v\delta(x) + 2v\delta(\ell-x)}{4(\alpha \ell + 4v)} \label{pps} \\
P_{+-}(x) &= \frac{\alpha + 4v\delta(x)}{4(\alpha\ell + 4v)} \label{pms}
\end{align}
in the limit $L\to\infty$
\footnote{Here, the delta functions are to be thought of as slightly displaced into the interior region from the boundaries at $x=0,\ell$, so that the integral $\int_0^{\ell} {\rm d}x \delta(x) = 1$.}.
In contrast to the other two limits considered so far, all three components of the stationary distribution survive in the scaling limit.  The extended and attractive components are present in the $++$ and $--$ sectors, Eq.~(\ref{pps}). In particular, the lengthscale $\xi \simeq 1/(2\omega)^{1/2}$ of the exponential decay corresponds to a microscopically large number of lattice sites of order $\sqrt{L}$. This is however small on the macroscopic scale, where each unit of length comprises $\sim L$ lattice sites: thus the attraction is confined to a fraction $\sim1/\sqrt{L}$ of the total system. At the same time, the amplitude of this exponential decay diverges as $\sqrt{L}$, and hence this component is manifested as the delta function appearing in (\ref{pps})---this delta function thus represents an attractive state in which particles move together with zero separation.  Meanwhile, the extended and jammed components appear in (\ref{pms}), where here the delta function has its origins in the Kronecker delta that appears in (\ref{pm}) and represents a jammed configuration.  From (\ref{pms}), we see that the particles spend a fraction $v / (\alpha\ell + 4v)$ of time in each of the two symmetrically-related jammed configurations and from (\ref{pps}) that a fraction $v/(2(\alpha\ell + 4v))$ is spent in each of the four attractive states with zero separation. Thus the total fraction of time spent in a state  in which particles are adjacent ($x=0$ or $\ell$) is $4v/(\alpha\ell + 4v)$.

We can also determine some features of the dynamics in the scaling limit.  In particular, the mean time spent in an adjacent state after a collision can be worked out from the fact that this state is left after exactly $2k$ velocity reversal events ($k=1,2,\ldots$) with probability $2^{-k}$. This is because particles are necessarily in the jammed state when they collide: the first reversal always causes the particles to both move at speed $v$ whilst remaining adjacent, and the next reversal either causes the particles to move apart or to re-enter a jammed configuration, each with equal probability. Since the total velocity reversal rate is $2\omega=\alpha$, it follows that the mean time between reversals is $1/\alpha$, and the mean time spent in an adjacent state is $4/\alpha$. Comparing this with the above result for the total fraction of the time spent in such a state, we deduce that the mean time between entering and leaving a non-adjacent state is $\ell/v$ (a result confirmed with an explicit first-passage time calculation in Appendix~\ref{app:fpt}). Intriguingly, this result is independent of the tumble rate $\alpha$, despite the fact that particles must typically tumble over this lifetime: otherwise, the time spent in this state would be close to its minimum value  $\ell/2v$. 

We now outline the derivation of Eqs.~(\ref{pp}) and (\ref{pm}). We start with the set of master equations governing the time evolution of the probabilities in the four velocity sectors:
\begin{align}
\dot{P}_{++}(n) &= P_{++}(n-1)I_{n>1} + P_{++}(n+1)I_{L-n>1}  \nonumber\\
&\phantom{=} + \omega[P_{+-}(n) + P_{-+}(n) ] \nonumber\\
&\phantom{=} - P_{++}(n) [2\omega + I_{n>1} + I_{L-n>1}] \label{ME pp} \\
\dot{P}_{+-}(n) &= 2P_{+-}(n+1)I_{L-n>1} + \omega[P_{++}(n) + P_{--}(n) ] \nonumber\\
&\phantom{=} - P_{+-}(n)[2\omega + 2I_{n>1}] \label{ME pm} 
\end{align}
along with  counterparts for $P_{-+}(n)$ and $P_{--}(n)$ which follow from the symmetries $P_{--}(n)=P_{++}(n)$ and $P_{-+}(n) = P_{+-}(L-n)$. In these equations the indicator $I_{k>1} = 1$ if $k>1$ and is zero otherwise. One can, of course,  check that Eqs.~(\ref{pp}) and (\ref{pm}), supplemented with (\ref{z})--(\ref{Z}), give the stationary solution of these equations. To actually construct the stationary solution, we introduce the generating functions $G_{\sigma_1\sigma_2}(x) = \sum_{n=1}^{L-1}x^n P_{\sigma_1\sigma_2}(n)$. Packaging these generating functions into a vector $\vec{G}(x)$, and performing the appropriate summations, we obtain a linear system $A(x) \vec{G}(x) = \vec{b}(x)$ where the elements of $\vec{b}$ do not involve the functions $G_{\sigma_1\sigma_2}(x)$. Then, it remains to evaluate  $\vec{G}(x) = A^{-1}(x) \vec{b}(x)$.  

In order to obtain (\ref{pp}) and (\ref{pm}) one must invert $\vec{G}(x)$. However, one still needs to fix $P_{++}(1)$ and $P_{+-}(1)$ , which are not \emph{a priori} known. These constants are fixed by noting that $A^{-1}(x)$ has poles at $x=1, x=z$ and $x=1/z$, where $z$ and $1/z$ are the two roots of the symmetric polynomial $x^{2} - 2(1+\omega)x + 1 = 0$. This implies an apparent divergence in the generating functions $G_{\sigma_{1}\sigma_{2}}(x)$ which is inconsistent with the fact that these functions are polynomials of degree $L-1$ and finite for all $x$. Therefore, the poles in $A^{-1}(x)$ must be canceled by zeros in the numerator $\vec{b}(x)$. This nontrivial pole-zero cancelation implies one relation between the two constants, $P_{++}(1)$ and $P_{+-}(1)$. The other required relation is given by the normalization of  probability.
Fixing the constants (see Appendix~\ref{app:gf} for details), one finds that the poles of $A^{-1}(x)$ at $x=1,z$ and $1/z$ correspond to a constant term and terms in $z^n$ and $z^{-n}$ in $P_{\sigma_1\sigma_2}(n)$, respectively, as in Eqs.~(\ref{pp}) and (\ref{pm}). 

\begin{figure}[h]
  \centering
    \includegraphics[width=\linewidth]{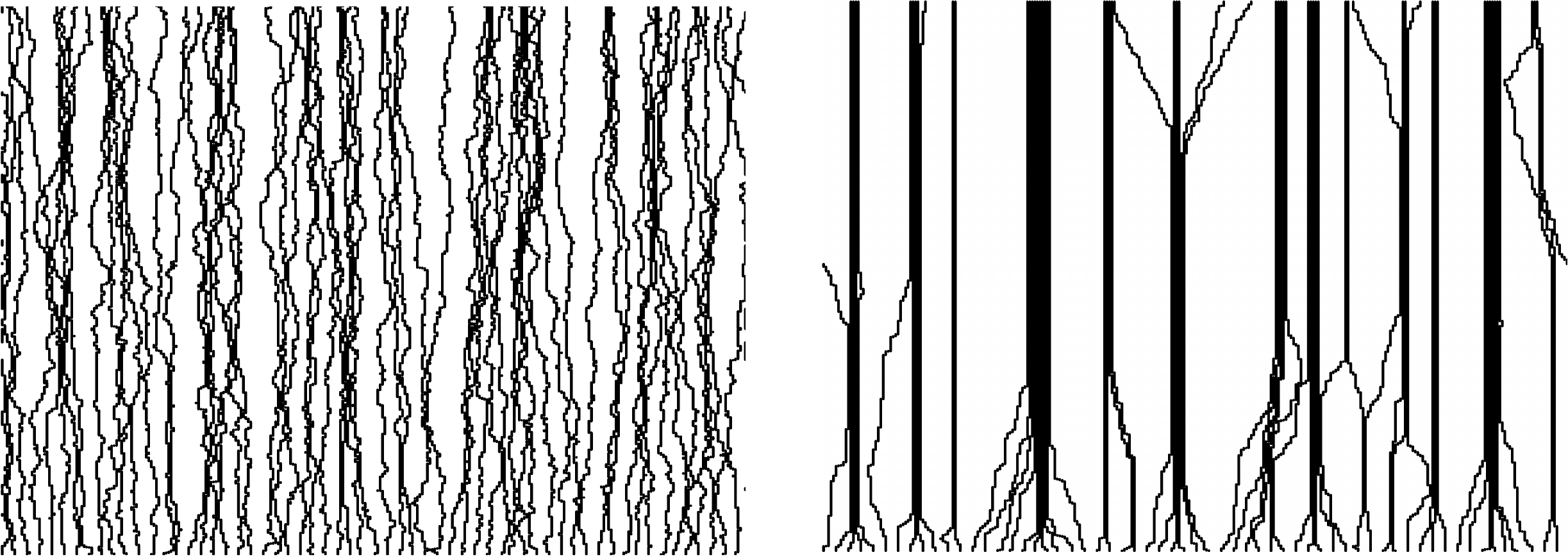}
    \caption{Space-time plots (time in the vertical direction) of 60 hard-core particles undergoing symmetric random walks (left) and run-and-tumble motion (right) on a lattice of $300$ sites. The initial condition and the particle hop rate is the same in both cases. In the run-and-tumble dynamics, $\omega=0.01$.  The clustering of particles induced by the nonequilibrium run-and-tumble dynamics is clearly evident (see also \cite{Thompson2011,Soto2014}).}
    \label{fig:sim}
\end{figure}

We conclude by considering how knowledge of an exact pair potential may bear on generalizations to many-body and higher-dimensional systems. We discuss the many-body case first. In Fig.~\ref{fig:sim} we compare simulations of hard-core particles in one dimension that hop with equal probability to the left or the right in each timestep (left panel) with the run-and-tumble dynamics that is the focus of this work (right panel). The former dynamics is a diffusion process satisfying detailed balance, which relaxes to a homogeneous steady state where all configurations are equally likely. Strikingly, breaking detailed balance by introducing run-and-tumble dynamics causes an inhomogeneous steady state with multiple clusters to appear (see also \cite{Thompson2011,Soto2014}).  Our calculations suggest that the jamming and attraction of pairs of particles may be responsible for this effect.  An important open question is whether an effective many-body potential, obtained by treating a summation of the pair potentials presented here as an ansatz, correctly predicts the physics of the many-body state.

In principle, it should also be possible to generalize the exact calculation of the pair potential to two or more dimensions. In particular, in two dimensions, a pair of diffusing particles (of finite size) will eventually collide with each other \cite{Redner2001}. By analogy with the one-dimensional problem here, we expect the jammed state to be present for a finite fraction of the time, and for a jammed state `echo' to give rise to an attractive interaction. The form of the corresponding potential might provide insight into whether one expects dense clusters to coarsen indefinitely leading to phase separation which, for active particles with direct repulsive interactions, experiments and simulations have been unable to observe directly due to the slow dynamics of aggregation \cite{Buttinoni2013,Stenhammar2013,Soto2014} .

Finally, the significance of the jamming and attraction established here in a simple model could be determined by investigating the effect of additional features of bacterial dynamics on the pair potential such as a finite tumbling duration \cite{Berg1972}, variable run velocity in response to chemical potentials \cite{deGennes2004,Kafri2008} or hydrodynamic interactions between particles \cite{Lauga2009}. Testing our predictions directly might be possible if micro-channel experiments confining a single run-and-tumble bacterium to one dimension whilst retaining its bulk motility pattern were extended to two interacting bacteria \cite{Mannik2009}.  Of course, the greatest insights of all would come from exact solutions of the many-body problem in arbitrary dimensions. This, however, remains a theoretical challenge.

\begin{acknowledgments}
The authors thank Davide Marenduzzo, Mike Cates and David Mukamel for detailed comments on the manuscript. ABS and MRE acknowledge support from EPSRC under a studentship and grant number EP/J007404/1.
\end{acknowledgments}

\appendix

\section{Derivation of the stationary distribution}
\label{app:gf}

In the main text, the master equations for $P_{++}(n)$ and $P_{--}(n)$ were given as
\begin{align}
\dot{P}_{++}(n) &= P_{++}(n-1)I_{n>1}+ P_{++}(n+1)I_{L-n>1}  \nonumber\\
&\phantom{=} + \omega[P_{+-}(n) + P_{-+}(n) ] \nonumber\\
&\phantom{=} - P_{++}(n) [2\omega + I_{n>1} + I_{L-n>1}] \\
\dot{P}_{+-}(n) &= 2P_{+-}(n+1)I_{L-n>1} + \omega[P_{++}(n) + P_{--}(n) ] \nonumber\\
&\phantom{=} - P_{+-}(n)[2\omega + 2I_{n>1}] \;.
\end{align}
Introducing the generating function $G_{\sigma_1\sigma_2}(x)=\sum_{n=1}^{L-1} x^n P_{\sigma_1\sigma_2}(n)$ we find
\begin{align}
\dot{G}_{++}(x) &= (x+x^{-1} - 2[1+\omega]) G_{++}(x) \nonumber\\
&\phantom{=} - (1-x)(1-x^{L-1})P_{++}(1) \nonumber\\
&\phantom{=} + \omega[ G_{+-}(x)+G_{-+}(x) ] \label{Gppdot}\\ 
\dot{G}_{+-}(x) &= 2(x^{-1} - [1+\omega]) G_{+-}(x) \nonumber\\
&\phantom{=} - 2(1-x)P_{+-}(1) \nonumber\\
&\phantom{=} + \omega[ G_{++}(x)+G_{--}(x) ] \label{Gpmdot} 
\end{align}
where we have used the fact that $P_{++}(L-1)=P_{++}(1)$. More generally, the symmetries $P_{++}(n)=P_{--}(n)=P_{++}(L-n)$ and $P_{+-}(n)=P_{-+}(L-n)$ in the stationary probability distribution translate to symmetries $G_{++}(x)=G_{--}(x)=x^L G_{++}(x^{-1})$ and $G_{+-}(x) = x^L G_{-+}(x^{-1})$ in the stationary values of their generating functions.  By exploiting these symmetries, we find from (\ref{Gppdot}) and (\ref{Gpmdot}) that the stationary generating functions must be the solution of the linear system
\begin{multline}
\left( \begin{array}{ccc}
\mu(x) + \nu(x) & \omega & \omega \\
\omega & \nu(x) & 0 \\
\omega & 0 & \mu(x)
\end{array} \right)
\left( \begin{array}{c}
G_{++}(x) \\ G_{+-}(x) \\ G_{-+}(x)
\end{array} \right)
= \\
(1-x) \left( \begin{array}{c}
(1-x^{L-1})P_{++}(1) \\
P_{+-}(1) \\
-x^{L-1} P_{+-}(1)
\end{array} \right) 
\end{multline}
where $\mu(x) = x - (1+\omega)$ and $\nu(x) = x^{-1}-(1+\omega) = \mu(1/x)$.  The inverse of the matrix appearing on the left-hand side of this expression is
\begin{multline}
\frac{x^2}{(1+\omega)(x-z)(x-\frac{1}{z})(1-x)(x-1)} \times \\
\left( \begin{array}{ccc}
\mu\nu & -\mu\omega & -\nu\omega \\
-\mu\omega & \mu(\mu+\nu)-\omega^2 & \omega^2 \\
-\nu\omega & \omega^2 & \nu(\mu+\nu)-\omega^2
\end{array} \right) \;, 
\end{multline}
where $z$ and $1/z$ are the two roots of $x^2 -2(1+\omega)x +1$ and recalling that $\mu$ and $\nu$ are functions of $x$.

This yields, for example,
\begin{multline}
G_{++}(x) = \frac{x^2}{(1+\omega)(x-1)(x-z)(x-\frac{1}{z})} \times \\
\big\{ \mu(x)\nu(x)[1-x^{L-1}] P_{++}(1)  +\\ \omega  [\nu(x) x^{L-1} -  \mu(x) ] P_{+-}(1) \big\} \;. 
\end{multline}
For general values of $P_{++}(1)$ and $P_{+-}(1)$, this gives an infinite series in $x$. However, it must terminate at order $x^{L-1}$, due to the original definition of the generating function. In particular, this implies that $G_{++}(x)$ should not diverge in the limits $x\to1$, $x\to z$ or $x\to 1/z$. Since $\nu(1)=\mu(1)$, we already have that the $x=1$ pole is canceled by a zero in the numerator. For this also to be the case at $x=z$, we must have
\begin{multline}
\mu(z)\nu(z) P_{++}(1) - \omega \mu(z) P_{+-}(1) =\\ z^{L-1} \left[ \mu(z)\nu(z) P_{++}(1) - \omega \nu(z) P_{+-}(1) \right] \;. 
\label{vanish}
\end{multline}
Since $\mu(x)=\nu(1/x)$, we find that the pole at $x=1/z$ is also canceled if this relation holds. Furthermore, using the fact that $\mu(z) = - \nu(z) = \frac{1}{2}(z - 1/z)$ and that $\omega = (z-1)^2/2z$, one can determine that
\begin{equation}
\frac{P_{++}(1)}{P_{+-}(1)} = \frac{1+z^{L-1}}{1-z^{L-1}} \frac{(1-z)^2}{1-z^2} \;. 
\label{ratio}
\end{equation}

To actually invert the generating function, we first note that
\begin{equation}
G_{++}(x) =
 \frac{x}{1+\omega}  \frac{x J(x)}{(x-1)(x-z)(x-\frac{1}{z})} 
 + x^L H_{++}(x) 
\label{Gpprelax}
\end{equation}
where 
\begin{equation}
J(x) =  \mu(x) \nu(x) P_{++}(1) - \omega \mu(x) P_{+-}(1) 
\end{equation}
and $H_{++}(x)$ is some power series in $x$. Now $P_{++}(n)$ is given by the coefficient of the $x^n$ term in $G_{++}(x)$. Since $n<L$, it follows that none of the terms in $H_{++}(x)$ contribute to the $P_{++}(n)$ of interest. In other words, to access $P_{++}(n)$, we read off the coefficient of $x^n$ in the first term of (\ref{Gpprelax}). Since the combination $xJ(x)$ is quadratic, and the denominator is cubic, we can apply a partial fraction decomposition to find
\begin{multline}
G_{++}(x) = \frac{x}{1+\omega} \bigg[ \frac{J(1)}{(1-z)(\frac{1}{z}-1)(1-x)} + \\
\frac{1}{(z-1)(\frac{1}{z}-z)} \left(  \frac{z J(1/z)}{1-zx} + \frac{J(z)}{1-x/z} \right)  \bigg] + x^L H_{++}(x) 
\label{parfracJ}
\end{multline}
Now, from (\ref{vanish}) and the fact that $\mu(x)=\nu(1/x)$ it follows that $J(z) = z^{L-1} J(1/z)$.  This implies that we can write
\begin{multline}
G_{++}(x) = \frac{x}{1+\omega} \bigg[ \frac{J(1)}{(1-z)(\frac{1}{z}-1)(1-x)} + \\
\frac{J(1/z)}{(z-1)(\frac{1}{z}-z)} \left(  \frac{z}{1-zx} + \frac{z^{L-1}}{1-x/z} \right)  \bigg] + x^L H_{++}(x) 
\end{multline}
which has the structure
\begin{equation}
G_{++}(x) = x\bigg[ \frac{a(z)}{1-x} +  \frac{b(z) z}{1-zx} + \frac{b(z) z^{L-1}}{1-x/z}  \bigg] + x^L H_{++}(x) 
\label{Gppdecomp}
\end{equation}
where
\begin{align}
a(z) &= \frac{J(1)}{(1+\omega)(\frac{1}{z}-1)(1-z)} \\
     &= \frac{(1-z)^2(1-z^L)}{(1+z)(1+z^2)(1-z^{L-1})} P_{+-}(1)\\ 
\intertext{and}
b(z) &= \frac{J(1/z)}{(1+\omega)(z-1)(\frac{1}{z}-z)} \\
     &= \frac{1-z}{(1+z^2)(1-z^{L-1})} P_{+-}(1) \;. 
\end{align}
Here we have used (\ref{ratio}) to express everything in terms of a single unknown constant $P_{+-}(1)$ that will be fixed by normalization.  Taking
\begin{equation}
\label{fixnorm}
P_{+-}(1) = \frac{(1+z)(1+z^2)(1-z^{L-1})}{Z} 
\end{equation}
and reading off the coefficient of $z^n$ in (\ref{Gppdecomp}) yields
\begin{equation}
P_{++}(n) = \frac{1}{Z}\left[ q(z) + p(z) (z^{n} + z^{L-n}) \right]  
\end{equation}
where the functions $p(z) = Z b(z)$ and $q(z) = Z a(z)$ have the functional forms that are given in the main text. 

The inversion of $G_{+-}(x)$ proceeds similarly, with a subtlety arising from the jamming occurring in this sector. Here we find that
\begin{multline}
G_{+-}(x) = - \frac{x}{1+\omega} \frac{xK(x)}{(x-1)(x-z)(x-\frac{1}{z})} + \\
\frac{x \mu(x) P_{+-}(1)}{(1+\omega)(x-1)} + x^L H_{+-}(x)
\end{multline}
where
\begin{equation}
K(x) = \omega [ \mu(x) P_{++}(1) + \omega P_{+-}(1) ] \;,
\end{equation}
and again the combination $xK(x)$ is quadratic in $x$.  The additional term that appears in the generating function would be cubic in $x$ if it were brought over a common denominator: this would not then be amenable to a partial fraction decomposition. The significance of the extra term is that it can ascribe an anomalously large weight to the jammed configurate.

To see this we perform the partial fraction decomposition, which gives analogously to (\ref{parfracJ}), 
\begin{widetext}
\begin{multline}
G_{+-}(x) = \frac{x}{1+\omega} \bigg[ \bigg( (1+\omega) P_{+-}(1) -
\frac{K(1)}{(1-z)(\frac{1}{z}-1)} - x P_{+-}(1) \bigg) \frac{1}{1-x}  + \\
\frac{1}{(1-z)(\frac{1}{z}-z)} \left(  \frac{z K(1/z)}{1-zx} + \frac{K(z)}{1-x/z}  \right)  \bigg] + x^L H_{+-}(x) \;. 
\end{multline}
\end{widetext}
Using (\ref{vanish}), along with the facts that $\mu(x)=\nu(1/x)$ and $\mu(z)=-\nu(z)$, one can show that $K(z) = -z^{L-1} K(1/z)$. This implies that $G_{+-}(x)$  has the structure
\begin{multline}
\label{Gpmdecomp}
G_{+-}(x) = \\
x\bigg[ \frac{a'(z) + c'(z) - x c'(z)}{1-x}  + \frac{b'(z) z}{1-zx} - \frac{b'(z) z^{L-1}}{1-x/z}\bigg]\\  + x^L H_{++}(x) 
\end{multline}
where
\begin{align}
a'(z) &= P_{+-}(1) - \frac{K(1)}{(1+\omega)(1-z)(\frac{1}{z}-1)} - \frac{P_{+-}(1)}{1+\omega} \\
      &= a(z) \\
b'(z) &= \frac{1}{1+\omega} \frac{K(1/z)}{(1-z)(\frac{1}{z}-z)} \\
      &= \frac{(1-z)^2}{(1+z)(1+z^2)(1-z^{L-1})} P_{+-}(1) \\
c'(z) &= \frac{P_{+-}(1)}{1+\omega} = \frac{2z}{1+z^2} P_{+-}(1) \;.
\end{align}
Substituting in (\ref{fixnorm}) and reading off the coefficient of $x^n$ in (\ref{Gpmdecomp}) finally yields
\begin{align}
P_{+-}(n) &= \frac{1}{Z} \left[ p'(z) (z^n - z^{L-n}) + q(z) + \delta_{n,1} \Delta(z) \right]
\end{align}
where $p'(z) = Z b'(z)$, $q(z) = Z a(z)$ and $\Delta(z) = Z c'(z)$ once again have the functional forms that are given in the main text. 
Expressions for $P_{--}(n)$ and $P_{-+}(n)$ follow from the symmetries $P_{--}(n)=P_{++}(n)$ and $P_{-+}(n) = P_{+-}(L-n)$.

\section{First-passage time calculation of the mean time spent in a nonadjacant state}
\label{app:fpt}

Our aim here is to calculate the mean time $\bar{T}_d$ that the particles spend in a nonadjacant state (i.e., one with separation $x>0$) after it is entered. Here we work directly in the scaling limit of the model where particles move ballistically with velocity $v$ and tumble at rate $\alpha$. Recall that tumbling leads to a velocity reversal with probability $\frac{1}{2}$. More generally we are interested in the mean first-passage time $\bar{T}_{A}(x)$ for two particles that are approaching with separation $x$ and closing speed $2v$ to reach the state of zero separation. Then, $\bar{T}_d = \bar{T}_{A}(L)$.

A differential equation that this quantity must satisfy is obtained by considering what happens in a short time interval $\delta t$ when the particles are approaching with separation $x$. With probability ${\rm e}^{-\alpha\delta t}$ the particles run without either reversing its velocity. If this happens, the particle separation decreases to $x-2v\delta t$. Alternatively, a velocity reversal occurs at a time $\delta t'$ which is distributed as $\alpha {\rm e}^{-\alpha \delta t'}$. When this happens, a state in which the two particles are moving in the same direction with constant separation $x-2\tau\delta t'$.  By introducing the mean time for two particles following each other at separation $x$ to meet as $\bar{T}_{F}(x)$, we have
\begin{multline}
\bar{T}_{A}(x) ={\rm e}^{-\alpha\delta t}  \left[ \delta t + \bar{T}_{A}(x-2v\delta t) \right]  + \\\int_0^{\delta t} {\rm d}(\delta t') \alpha {\rm e}^{-\alpha \delta t'} \left[ \delta t' + \bar{T}_{F}(x-2v\delta t') \right] \;.
\end{multline}
Taylor expanding the right-hand side, and dropping terms of order $(\delta t)^2$ and higher, we find
\begin{multline}
\bar{T}_{A}(x) = \bar{T}_{A}(x) + \\
\left[ 1 - 2v \frac{\rm d}{{\rm d} x} \bar{T}_A(x) + \alpha \left( \bar{T}_F(x) -\bar{T}_A(x) \right) \right] \delta t \label{tax} \;.
\end{multline}
Now, while particles are following each other at the same velocity, they maintain their separation until one of them reverses its velocity. The mean time until this happens is $\frac{1}{\alpha}$, at which time the particles with equal probability either start approaching each other  at separation $x$ or start receding from each other at separation $x$. By symmetry, this latter state is the same as approaching at separation $L-x$, and so
\begin{equation}
\label{tfx}
\bar{T}_{F}(x) = \frac{1}{2\tau} + \frac{1}{2} \left[ \bar{T}_{A}(x) + \bar{T}_A(L-x) \right] \;.
\end{equation}
Now, substituting this into (\ref{tax}) we arrive at 
\begin{align}
\frac{\rm d}{{\rm d} x} \bar{T}_{A}(x) &=\frac{1}{v} + \frac{\alpha}{4v} \left[ \bar{T}_{A}(L-x) - \bar{T}_{A}(x) \right] \;.
\label{deq}
\end{align}

To solve this equation we introduce the decomposition
\begin{equation}
\bar{T}_A(x) = F(x) + G(x)
\end{equation}
where $F(L-x)=F(x)$ and $G(L-x)=-G(x)$.  That is,
\begin{align}
F(x) &= \frac{1}{2} \left[ \bar{T}_A(x) + \bar{T}_A(L-x) \right] \\
G(x) &= \frac{1}{2} \left[ \bar{T}_A(x) - \bar{T}_A(L-x) \right] \;.
\end{align}
Then
\begin{align}
\frac{{\rm d} F}{{\rm d} x} &= \frac{1}{2} \left[ \frac{\rm d}{{\rm d} x} \bar{T}_A(x) - \frac{\rm d}{{\rm d} x} \bar{T}_A(L-x) \right] = - \frac{\alpha}{2v} G(x) \\
\frac{{\rm d} G}{{\rm d} x} &= \frac{1}{2} \left[ \frac{\rm d}{{\rm d} x} \bar{T}_A(x) + \frac{\rm d}{{\rm d} x} \bar{T}_A(L-x) \right] = \frac{1}{v} \;.
\end{align}
Hence,
\begin{align}
G(x) &= G_0 + \frac{x}{v} \\ \implies F(x) &= F_0 - \frac{\tau G_0 x}{v} - \frac{\alpha x^2}{4v^2} \;.
\end{align}
The constant $G_0$ is fixed by the symmetry $G(L-x)=-G(x)$, that is
\begin{equation}
G_0 + \frac{L-x}{v} = - G_0 - \frac{x}{v} \quad\implies\quad
G_0 = -\frac{L}{2v} \;.
\end{equation}
We then find that $F(x)=F(L-x)$ for any value of the constant $F_0$, since
\begin{equation}
F(x) = F_0 + \frac{\alpha (L-x)}{4v^2} \;.
\end{equation}
The remaining constant $F_0$ is fixed using the boundary condition at zero separation, $\bar{T}_{A}(0)=0$. This implies that
\begin{equation}
\bar{T}_A(x) = \frac{x}{v} + \frac{\alpha x(L-x)}{4v^2} \;.
\end{equation}
In particular, mean time between entering and exiting a nonadjacent state is
\begin{equation}
\bar{T}_d = \bar{T}_A(L) = \frac{L}{v} \;,
\end{equation}
thereby confirming the tumble-rate independence that was obtained by alternative means in the main text.

\bibliography{MyCollection}

\end{document}